%% file: art3.tex
\documentstyle[preprint,aps]{rewtex}
\begin{document}
\draft
\title{On the Point-Splitting Method \\ of the Commutator Anomaly \\
 of Gauss Law Operators}
\author{R.A. Bertlmann\thanks{e-mail address -- bertlman@pap.univie.ac.at}}
\address{Institut f\"ur Theoretische Physik,
Universit\"at Wien,
Boltzmanngasse 5, A-1090 Wien, Austria}
\author{Tom\'a\v s S\'ykora \thanks{e-mail address
-- sykora@hp03.troja.mff.cuni.cz}}
\address{Department of Nuclear Centre,
 Faculty of Mathematics and Physics, Charles University,\\ V
Hole\v sovi\v ck\'ach 2, 180 00 Prague, Czech Republik}
\date{July 26, 1996}
\maketitle
\begin{abstract}
We analyze the generalized point-splitting method and  Jo's result
for the commutator anomaly.
We find that certain classes of general regularization kernels satisfying
 integral conditions provide a {\sl unique} result, which, however, {\sl differs}
from Faddeev's cohomological result.
\end{abstract}

\section{Introduction}
\label{I}
\renewcommand{\theequation}{1.\arabic{equation}}
There is now an elegant cohomological theory -- the socalled Stora-Zumino chain of descent
equations \cite{1,2} -- established which describes the anomalies of quantum field theory
(for a recent overview see \cite{3,4}). The 1-cocycle is identified with the anomaly
in the covariant divergence of the non-Abelian chiral fermion current \cite{1,2},
the 2-cocycle with the anomalous term -- the Schwinger term \cite{5} -- in the commutator
of the gauge group generators occuring in the same anomalous theory \cite{6,7,8}
(for an overview see \cite{9,10}). It is this anomalous (equal time) commutator we are concerned with
\begin{equation}
i[{\cal G}^{a}({\bf x}),{\cal G}^{b}({\bf y})]=f^{abc}{\cal G}^{c}\delta^{3}({\bf x-y})+S^{ab}({\bf
x-y})\label{1.1}.
\end{equation}
The generator -- the Gauss-law operator -- consists of 2 parts
\begin{equation}
{\cal G}^{a}({\bf x}) = \delta^{a}({\bf x}) + \rho^{a}({\bf x}),
\end{equation}
the generator $\delta^{a}({\bf x})$ of gauge transformations for
the gauge potentials and the generator
$\rho^{a}({\bf x})$ of the gauge transformations for the fermionic fields
\begin{eqnarray}
\delta^{a}({\bf x})&=&-({\bf D\cdot E})^{a}({\bf x})=
i\,D_{i}^{ba}\,\frac{\delta}{\delta A_{i}^{b}}({\bf x}), \\
\rho^{a}({\bf x})&=&-i\:\psi^{\dagger}({\bf x})T^{a}
\frac{1-\gamma_{5}}{2}\psi({\bf x}).
\end{eqnarray}
$E^{a}_{i}$ is the non-Abelian electric field, $D_{i}^{ab}=
\delta^{ab}\partial_{i}+f^{abc}A^{c}_{i}$ the covariant deriva\-tive; the group
matrices
$T^{a}$ are
anti-Hermitian satisfying
\begin{equation}
[T^{a},T^{b}]=f^{abc}T^{c},
\end{equation}
and finally $\gamma_{5}$ is chosen like
 $\gamma_{5}=i\gamma^{0}\gamma^{1}\gamma^{2}\gamma^{3}$.

The solution for this additional anomalous term in the commutator -- which causes difficulties
when quantizing the theory -- has been found by Faddeev \cite{7} on a cohomological basis
\begin{equation}
S^{ab}({\bf x,y})=-\frac{i}{24\pi^{2}}\varepsilon^{ijk}\mbox{tr}\{T^{a},T^{b}\}
\partial_{i}A_{j}\partial_{k}\delta^{3}({\bf x-y})\label{1.6}.
\end{equation}
This cohomological result has been verified by computing the commutator with
the Bjorken-Johnson-Low
procedure \cite{11,12,13,14}, or by working with geometric methods
\cite{15,16,17,18,19,20,21}. However, as Jo \cite{11} discovered a generalized point-splitting
method where the time is fixed {\sl does not} provide Faddeev's cohomological
re\-sult (\ref{1.6}), contrary
to claims in the literature \cite{22}. Furthermore, Jo located an inherent ambiguity in the procedure
due to the specific choice of the regularization kernels. (Note that we consider here the case
 of 1+3 dimensions, in 1+1 dimensions there occur no problems and all methods agree).
 Re-investigating the procedure we clarify this ambiguity and show how to overcome this
 problem. In fact, we find that a whole class of regularization kernels satisfying an
 integral condition provides a unique result.

\section { Generalized point-splitting method}
\renewcommand{\theequation}{2.\arabic{equation}}
\setcounter{equation}{0}
In order to define an operator
\begin{equation}
{\cal J}(f)=-i\int{d^{3}x\:{\psi^{\dagger}({\bf x})}f({\bf x})\frac{1-\gamma_{5}}{2}\psi({\bf x})},
\end{equation}
which has a singular behavior, we introduce a family of the smooth kernels
\begin{equation}
F({\bf x,y})=f\Bigl(\frac{\bf x+y}{2}\Bigr)\:f_{\mu_{f}}(|{\bf x-y}|),
\end{equation}
where
\begin{equation}
\lim_{\mu_{f}\to 0}f_{\mu_{f}}(|{\bf x-y}|)=\delta^{3}({\bf x-y}).
\end{equation}
The limit is understood in a distributional sense, so the
$f_{\mu_{f}}(|{\bf x-y}|)$ are $\delta$-like functions and the $f\Bigl(\frac{\bf x+y}{2}\Bigr)$
 contain matrices of the internal symmetry space.
For each such kernel $F({\bf x,y})$ we define the operator
\begin{equation}
{\cal J}(F)=-i\int{d^{3}x d^{3}y\:{\psi^{\dagger}}({\bf x})F({\bf x,y})\frac{1-\gamma_{5}}{2}\psi({\bf y})}.
\end{equation}
We also need
the Fourier transformations
\begin{eqnarray}
\tilde{F}({\bf p',p}) &=& \int{
d^{3}x d^{3}y\:
e^{i{\bf p'\cdot (x+y)}/2}
e^{i{\bf p\cdot (x-y)}}
F({\bf x,y})}, \nonumber \\
&=&\tilde{f}({\bf p'})\:\tilde{f}_{\mu_{f}}(|{\bf
p}|), \\
\tilde{f}({\bf p'}) &=&\int{d^{3}x\: e^{i{\bf p'\cdot x}}\:f({\bf
x})},\\
\tilde{f}_{\mu_{f}}(|{\bf p}|)&=&\int{d^{3}x\: e^{i{\bf p\cdot x}}\:f_{\mu_{f}}(|{\bf x}|)}.
\end{eqnarray}
The $\tilde{F}({\bf p',p})$ has the local limit
\begin{equation}
\lim_{\mu_{f}\to 0}\tilde{F}({\bf p',p})=\tilde{f}({\bf p'}),
\end{equation}
since
\begin{equation}
\lim_{\mu_{f}\to 0}\tilde{f}_{\mu_{f}}(|{\bf p}|)=1.
\end{equation}
These smeared operators ${\cal J}(F)$ are well defined in a Hilbert space
and satisfy the familiar commutation relations
\begin{equation}
i[{\cal J}(F), {\cal J}(G)]={\cal J}([F,G]),
\end{equation}
where the commutator $[F,G]$ means
\begin{equation}
[F,G]({\bf x,y})=\int{d^{3}z\:[F({\bf x,z})G({\bf z,y})-G({\bf x,z})F({\bf z,y})]}.
\end{equation}
However, in order to be able to perform the local limit we have to subtract the fixed-time vacuum
expectation value (VEV) of ${\cal J}(F)$
\begin{equation}
{\big\langle} {\cal J}(F){\big\rangle}_{A}= \int{
d^{3}x d^{3}y\:
\mbox{tr}P({\bf x,y}) F({\bf y,x})}=\mbox{Tr}FP,
\end{equation}
where
\begin{equation}
P({\bf x,y})=i\frac{1-\gamma_{5}}{2}{\big\langle}\psi({\bf x}){\psi^{\dagger}}
({\bf y}){\big\rangle}_{A}.
\end{equation}
In the local limit we need $P({\bf x,y})$ for ${\bf x}\approx{\bf y}$, which
diverges for
${\bf x}\rightarrow{\bf y}$. We extract
$P^{inf}= P({\bf x,y})_{\bf x\rightarrow\bf y}$ so that
$P-P^{inf}$ has a local limit.
Then we obtain the well-defined operator ${\cal J}(f)$  from the local limit of
a such regularized quantity
\begin{equation}
{\cal J}(f)=\lim_{\mu_{f} \to 0} {\cal J}_{reg}(F),
\end{equation}
with
\begin{equation}
{\cal J}_{reg}(F)={\cal J}(F)-\mbox{Tr}FP^{inf}.
\end{equation}
On the other hand, we also need an operator ${\cal T}(f)$ defined by
\begin{equation}
{\cal T}(f)=\int{d^{3}x\:f^{a}({\bf x})\delta^{a}}({\bf
x})=-i\int{d^{3}x\Big[{D_{i}f({\bf x})}\Bigr]^{b}
\frac{\delta}{\delta A_{i}^{b}({\bf x})}}.
\end{equation}
Now, in order to investigate the commutator (\ref{1.1}) we have to consider
\begin{equation}
i[{\cal T}(f)+{\cal J}_{reg}(F),{\cal T}(g)+{\cal J}_{reg}(G)]=
{\cal T}([f,g])+{\cal J}_{reg}([F,G])+S(F,G),
\end{equation}
and we have to compute the Schwinger term in the local limit
\begin{equation}
S(F,G)= i[{\cal T}(f),{\cal J}_{reg}(G)]+i[{\cal J}_{reg}(F),{\cal T}(g)]+
{\cal J}\bigl([F,G]\bigr)-{\cal J}_{reg}\bigl([F,G]\bigr),
\end{equation}
\begin{equation}
S(f,g)=\lim_{\mu_{f},\mu_{g} \to 0}S(F,G).
\end{equation}

\section{Jo'\lowercase{s} result for the commutator anomaly}
\label{3}
\renewcommand{\theequation}{3.\arabic{equation}}
\setcounter{equation}{0}
For symmetric regularization kernels the following commutators vanish
\begin{equation}
[{\cal T}(f),{\cal J}_{reg}(G)]=[{\cal J}_{reg}(F),{\cal T}(g)]\,=\,0
\end{equation}
and we have for the Schwinger term \cite{11}:
\begin{eqnarray}
S(F,G)&=&\mbox{Tr}[F,G]P^{inf} \nonumber \\
&=&\frac{1}{2}\int{\frac{d^{3}q}{(2\pi)^{3}}\:\mbox{tr}\:B^{i}({\bf q})
\int{\frac{d^{3}p'}{(2\pi)^{3}}\:[\chi^{i}(F,G;{\bf p',q})-\chi^{i}(G,F;{\bf
p',q})]}}, \label{3.3}
\end{eqnarray}
where $B^{i}({\bf q})$ is the Fourier transformation of
$B^{i}\big(\frac{\bf x+y}{2}\big)$ and
\begin{equation}
B^{i}({\bf x})=\varepsilon^{ijk}(\partial_{j}A_{k}+A_{j}A_{k})({\bf x}).
\end{equation}
The function
\begin{equation}
\chi^{i}(F,G;{\bf p',q})=\int{\frac{d^{3}p}{(2\pi)^{3}}
\:\frac{{p}^{i}}{|{\bf p}|^{3}}\:\tilde{F}\Bigl({\bf p',p}+\frac{\bf
q}{2}+\frac{\bf p'}{2}\Bigr)\tilde{G}
\Bigl({\bf -p'-q},{\bf p}+\frac{\bf p'}{2}\Bigr)}
\end{equation}
after expanding $\tilde{F}$ and $\tilde{G}$ can be rewritten as
\begin{eqnarray}
\chi^{i}(F,G;{\bf p',q})&=&\int{\frac{d^{3}p}{(2\pi)^{3}}
\:\frac{{p}^{i}}{|{\bf p}|^{3}}\Bigl[
\tilde{F}({\bf p',p})\tilde{G}({\bf -p'-q,p})\Bigr]}+\nonumber \\ \nonumber\\
&+&\frac{{ p'}^{j}}{2}\:\int{\frac{d^{3}p}{(2\pi)^{3}}
\:\frac{{p}^{i}}{|{\bf p}|^{3}}\frac{\partial}{\partial
p^{j}}\Bigl[
\tilde{F}({\bf p',p})\tilde{G}({\bf -p'-q,p})\Bigr]}+\nonumber \\ \nonumber\\
&+&\;\frac{{ q}^{j}}{2}
\:\int{\frac{d^{3}p}{(2\pi)^{3}}\:\frac{{p}^{i}}{|{\bf p}|^{3}}\Bigl[
\frac{\partial}{\partial
p^{j}}\tilde{F}({\bf p',p})\Bigr]\tilde{G}({\bf -p'-q,p})}+\nonumber \\ \nonumber\\
&+& \mbox{ higher-order derivative terms}.
\end{eqnarray}
The first integral is zero because the integrand is antisymetric under the change of ${\bf p}
\rightarrow -{\bf p}$.
The higher-order derivative terms vanish after the local limit.\footnote{
This is valid for all renormalization kernels
and not only for the
Gaussian ones used by Jo \cite{11}. }

Then the function $\chi^{i}$ can be separated into 2 parts
\begin{equation}
\chi^{i}(F,G;{\bf p',q})=\chi_{1}^{i}(F,G;{\bf p',q})+
\chi_{2}^{i}(F,G;{\bf p',q}),
\end{equation}
where
\begin{eqnarray}
\chi_{1}^{i}(F,G;{\bf p',q})&=&
\frac{{p'}^{j}}{2}\:\int{\frac{d^{3}p}{(2\pi)^{3}}
\:\frac{{p}^{i}}{|{\bf p}|^{3}}\frac{\partial}{\partial
p^{j}}\Bigl[
\tilde{F}({\bf p',p})\tilde{G}({\bf -p'-q,p})\Bigr]}, \\
\chi_{2}^{i}(F,G;{\bf p',q})&=&\frac{{ q}^{j}}{2}
\:\int{\frac{d^{3}p}{(2\pi)^{3}}\:\frac{{p}^{i}}{|{\bf p}|^{3}}\Bigl[
\frac{\partial}{\partial
p^{j}}\tilde{F}({\bf p',p})\Bigr]\tilde{G}({\bf -p'-q,p})}.
\end{eqnarray}
Whereas the first term is independent of the applied
regularization kernels -- the $\delta$-like functions $\tilde{f}_{\mu_{f}}(|{\bf
p}|)$,
$\tilde{g}_{\mu_{g}}(|{\bf p}|)$ -- providing such the unique result
\begin{equation}
\chi_{1}^{i}(F,G;{\bf p',q})=-\frac{{p'}^{i}}{12\pi^{2}}
\tilde{f}({\bf p'})\tilde{g}({\bf -p'-q})
\end{equation}
in the local limit $\mu_{f}$, $\mu_{g}\rightarrow 0$,
the second term is not.
It strongly depends on the kernels and for Jo's choice of Gaussian regularization kernels
\begin{eqnarray}
F({\bf x,y})&=&f\Bigl(\frac{\bf x+y}{2}\Bigr) \frac{1}{(4\pi\mu_{f})^{3/2}}\:e^{-({\bf x-y})^{2}/4\mu_{f}},\label{3.11}\\
G({\bf x,y})&=&g\Bigl(\frac{\bf x+y}{2}\Bigr) \frac{1}{(4\pi\mu_{g})^{3/2}}\:e^{-({\bf x-y})^{2}/4\mu_{g}},
\end{eqnarray}
or in momentum space
\begin{eqnarray}
\tilde{F}({\bf p',p})&=&\tilde{f}({\bf p'}) e^{-\mu_{f}{\bf p}^{2}},\\
\tilde{G}({\bf p',p})&=&\tilde{g}({\bf p'}) e^{-\mu_{g}{\bf p}^{2}},
\end{eqnarray}
the result is
\begin{equation}
\chi_{2}^{i}(F,G;{\bf p',q})=-\frac{1}{1+\mu}\frac{q^{i}}{12\pi^{2}}\tilde{f}({\bf p'})
\tilde{g}({\bf -p'-q}),
\end{equation}
where we have introduced the parameter $\mu\equiv\mu_{g}/\mu_{f}$.
Clearly, the local limit of $\chi_{2}^{i}$ depends on how $\mu_{f}$ and $\mu_{g}$ approach zero.
With this ambiguity, the final expression for the Schwinger term  becomes
$$
S(F,G)=\frac{-i}{24\pi^{2}}
\varepsilon^{ijk}\times
\;\;\;\;\;\;\;\;\;\;\;\;\;\;\;\;\;\;\;\;\;\;\;\;\;\;\;\;\;\;\;\;\;\;\;\;\;\;\;\;\;\;\;\;\;\;
\;\;\;\;\;\;\;
\;\;\;\;\;\;\;\;\;\;\;\;\;\;\;\;\;\;\;\;\;\;\;\;\;\;\;\;\;\;\;\;\;\;\;
$$
\begin{equation}
{\times}\int{
d^{3}x\:\mbox{tr}\Bigl[
(\partial_{j}A_{k}+A_{j}A_{k})
(\partial_{i}f\, g-\partial_{i}g\, f)+
\partial_{i}(A_{j}A_{k})\Bigl(\frac{1}{1+\mu}\,fg-
\frac{\mu}{1+\mu}\,gf\Bigr)\Bigr]}.
\end{equation}
As emphasized by Jo using different regularization kernels may give rise to
a different approach dependence, to a different dependence on $\mu$.
This is indeed the case as we shall demonstrate below.

\section{Power-like regularization kernels}
\label{4}
\begin{figure}
\caption{The combination of the Gaussian and power kernels, Eqs.
(\ref{4.10})
and (\ref{4.12}) $^{3}$, are plotted versus $\xi$ for the values of $b=3$ and $b=4$.}
\label{fig1}
\begin{center}
\input psbox.tex
$$\psboxscaled{700}{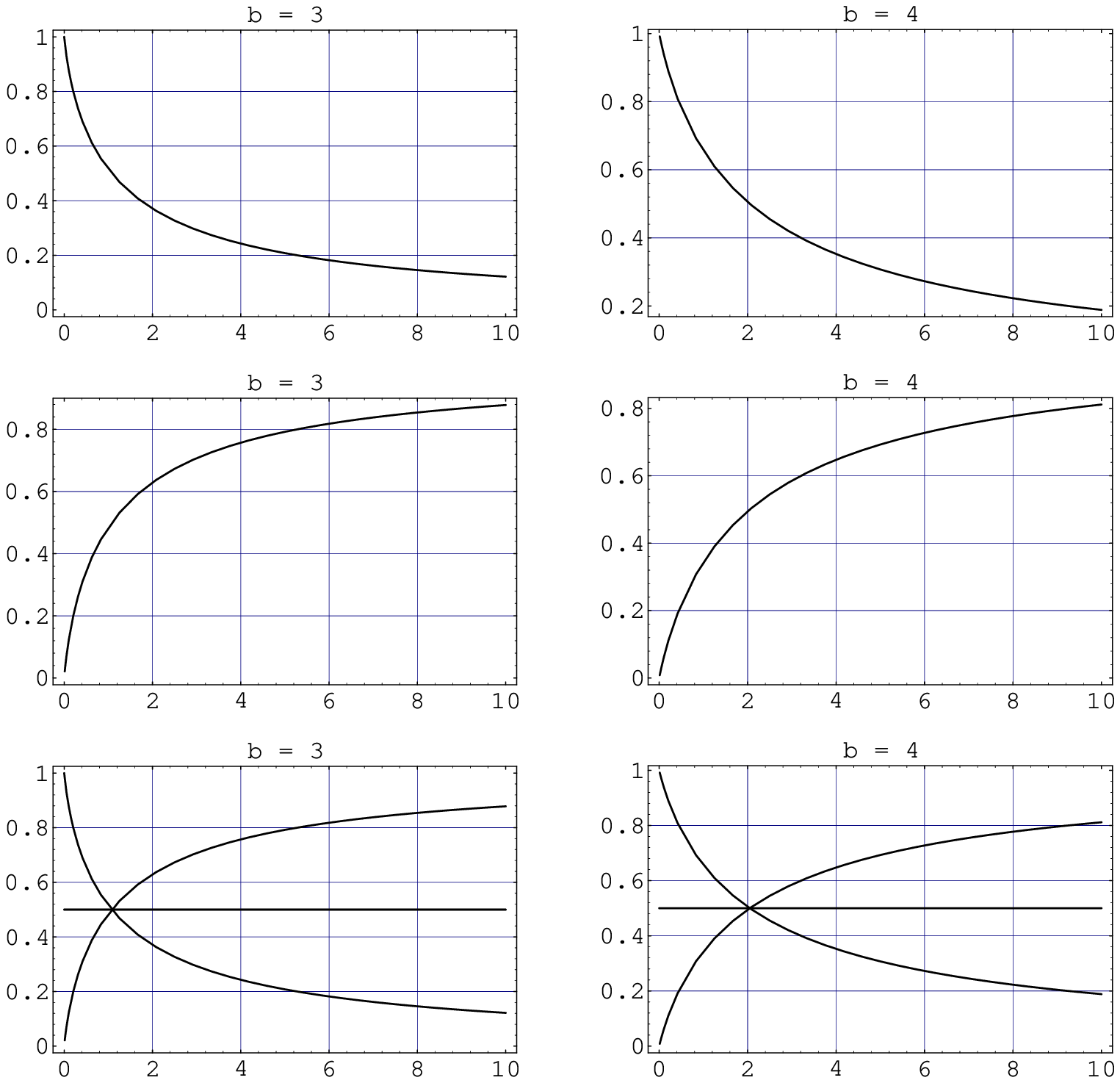}$$
\end{center}
\end{figure}

\renewcommand{\theequation}{4.\arabic{equation}}
\setcounter{equation}{0}
Let us choose a new set of $\delta$-like functions $\{f_{\mu_{f}}(|{\bf
x}|,b)\}$, the power functions \cite{23}
\begin{equation}
f_{\mu_{f}}(|{\bf x-y}|,b)=\frac{1}{N}\frac{\mu_{f}^{2b-3}}{(|{\bf x-y}|^{2}+\mu_{f}^{2})^{b}}\label{4.1},
\end{equation}
with the normalization (beta function)
\begin{equation}
N=2 \pi\, B(3/2,b-3/2)=2\pi \frac{\Gamma(\frac{3}{2})\Gamma(b-\frac{3}{2})}{\Gamma(b)}
\end{equation}
and $b\geq 3, b\in {\bf R}$. The
Fourier transforms are
\begin{equation}
\tilde{f}_{\mu_{f}}(|{\bf p}|,b)=\frac{1}{\tilde N}\,(\mu_{f}
|{\bf p}|)^{b-3/2}K_{b-3/2}(\mu_{f}|{\bf p}|),
\end{equation}
with
\begin{equation}
{\tilde N}=2^{b-5/2}\,\Gamma(b-3/2),
\end{equation}
and $K_{b-3/2}(\beta)$ is a Bessel function.
In this case we obtain for the ambiguous term \cite{24}
\begin{equation}
\chi_{2}^{i}(F,G;{\bf p',q})=-\frac{q^{i}}{12\pi^{2}}\tilde{f}({\bf p'})
\tilde{g}({\bf -p'-q})\frac{\mu^{2b-3}}{2}
F(2b-3,b-1/2;2b-2;1-\mu^{2})\label{4.5},
\end{equation}
where $F(a,b;c;z)$ denotes the hypergeometric function with the integral
representation (Re $c>$ Re $b>0$)
\begin{equation}
F(a,b;c;z)=\frac{1}{B(b,c-b)}\int_{0}^{1}{dt\; t^{b-1}(1-t)^{c-b-1}(1-zt)^{-a}}.
\end{equation}
Clearly, for the 2 parameter values\footnote{
For the case $\mu=0$ it is better to use the expression
\begin{displaymath}
\chi_{2}^{i}(F,G;{\bf p',q})=-\frac{q^{i}}{12\pi^{2}}\tilde{f}({\bf p'})
\tilde{g}({\bf -p'-q})\frac{\mu^{b-3/2}}{2^{2b-5}{\Gamma}^{2}(b-3/2)}\cdot
I(b,\mu),
\end{displaymath}
where
\begin{displaymath}
I(b,\mu)=\int _{0}^{\infty}{dt\: t^{2b-3} K_{b-5/2}(t) K_{b-3/2}(\mu t)}.
\end{displaymath} }
$\mu=0$ and $\mu\rightarrow\infty$ we recover -- for all values of $b$ -- Jo's
result (this must be the case for general reasons as we shall demonstrate
below). But also for $\mu=1$ the result (\ref{4.5}) agrees with Jo's result derived
with the Gaussian kernels.
Of course, for a general value of $\mu$ this is not so. For example, for $b=3$
we get
\begin{equation}
\chi_{2}^{i}(F,G;{\bf p',q})=-\frac{3\mu+1}{(1+\mu)^{3}}\frac{q^{i}}{12\pi^{2}}\tilde{f}{\bf(p'})\tilde{g}({\bf -p'-q}).
\end{equation}
Next we combine different $\delta$-like functions. For example, let us choose
the Gaussian kernel (\ref{3.11}) to regularize the operator ${\cal J}(f)$ and
the above
power kernel (\ref{4.1}) for ${\cal J}(g)$ then we obtain a different $\mu_{f}$,
$\mu_{g}$ dependence of the integral \cite{24}
\begin{equation}
\chi_{2}^{i}(F,G;{\bf p',q})=-\frac{q^{i}}{12\pi^{2}}
\tilde{f}{\bf(p'})\tilde{g}({\bf -p'-q})(b-\frac{3}{2})\xi^{b-3/2}U(b-1/2,b-1/2,\xi)\label{4.10},
\end{equation}
where $\xi\equiv \mu \mu_{g}/4$ and
$U(a,b,z)$ denotes the Whittaker function with integral representation
(Re $a>0$)
\begin{equation}
U(a,b,z)=\frac{1}{\Gamma (a)}\int_{0}^{\infty}{dt\; e^{-zt}t^{a-1}(1+t)^{b-a-1}}.
\end{equation}
If we interchange the kernels then we obtain again an other $\mu_{f}$,
$\mu_{g}$ dependence
\begin{equation}
\chi_{2}^{i}(F,G;{\bf p',q})=-\frac{q^{i}}{12\pi^{2}}
\tilde{f}{\bf(p'})\tilde{g}({\bf -p'-q})\xi^{b-3/2}U(b-3/2,b-3/2,\xi)\label{4.12}.
\end{equation}
The results (\ref{4.10}) and (\ref{4.12}) we have plotted\footnote{Up to the common factor
$-\frac{q^{i}}{12\pi^{2}}
\tilde{f}{\bf(p'})\tilde{g}({\bf -p'-q})$.}
on Fig.\ \ref{fig1} \cite{25}.
Again, for $\xi=0$ and
$\xi\rightarrow\infty$ we recover the previous cases but now the desired
agreement with the previous results, the value $1/2$ where both functions
(\ref{4.10}) and (\ref{4.12}) coincide$^{3}$, is given at different $\xi$ depending on the value
of $b$. This corresponds to taking a different limit procedure for each value of
$b$. The several $\xi$ values we have collected in Tab. \ref{tab1} \cite{25}. Of course, for
general values of $\xi$ the results differ from the previous ones.

So the above demonstrated dependence of the integral $\chi_{2}^{i}
(F,G;{\bf p',q})$ on the applied regularization kernels proves Jo's conjecture.

\section{Integral condition}
\label{5}
\renewcommand{\theequation}{5.\arabic{equation}}
\setcounter{equation}{0}
But we can overcome this ambiguity in a quite natural way. Let us consider
again the Schwinger term expression (\ref{3.3}) for general regularization kernels.
Since it is antisymmetric under interchange of $f$ and $g$ the final integral
in the term
\begin{equation}
\chi_{2}^{i}(F,G;{\bf p',q})=\frac{q^{i}}{12\pi^{2}}
\tilde{f}{\bf(p'})\tilde{g}({\bf -p'-q})
\lim_{\mu_{f},\mu_{g} \to 0}\int_{0}^{\infty}d|{\bf p}|\:\frac{\partial}{\partial |{\bf p}|}
{\tilde f}_{{\mu}_{f}}(|{\bf p}|)
\cdot{\tilde g}_{{\mu}_{g}}(|{\bf p}|)
\end{equation}
must be invariant under this interchange, so
$$
\lim_{\mu_{f},\mu_{g} \to 0}\int_{0}^{\infty}d|{\bf p}|\:\frac{\partial}{\partial |{\bf p}|}
{\tilde f}_{{\mu}_{f}}(|{\bf p}|)
\cdot{\tilde g}_{{\mu}_{g}}(|{\bf p}|)=\:\:\:\:\:\:\:\:\:\:\:\:\:\:\:\:\:\:\:\:\:
$$
\begin{equation}
\:\:\:\:\:\:\:\:\:\:\:\:\:\:\:\:\:\:=\lim_{\mu_{f},\mu_{g} \to 0}\int_{0}^{\infty}d|{\bf p}|\:
{\tilde f}_{{\mu}_{f}}(|{\bf p}|)\cdot\frac{\partial}
{\partial |{\bf p}|}{\tilde g}_{{\mu}_{g}}(|{\bf p}|)\label{5.2}.
\end{equation}
After partial integration follows
\begin{equation}
2\lim_{\mu_{f},\mu_{g} \to 0}\int_{0}^{\infty}d|{\bf p}|\:\frac{\partial}{\partial |{\bf p}|}
{\tilde f}_{{\mu}_{f}}(|{\bf p}|)
\cdot{\tilde g}_{{\mu}_{g}}(|{\bf p}|)=\lim_{\mu_{f},\mu_{g} \to 0}
\Bigl[{\tilde f}_{{\mu}_{f}}(|{\bf p}|)
\:{\tilde g}_{{\mu}_{g}}(|{\bf p}|)\Bigr]^{\infty}_{0}=-1,
\end{equation}
since $\delta$-like functions satisfy
\begin{equation}
{\tilde f}_{{\mu}_{f}}(\infty)={\tilde g}_{{\mu}_{g}}(\infty)=0 \mbox{\ \ \ \ \
and \ \ \ }
\lim_{\mu_{f} \to 0}{\tilde f}_{{\mu}_{f}}(0)=\lim_{\mu_{g} \to 0}{\tilde g}_{{\mu}_{g}}(0)=1.
\end{equation}
An other way of getting the condition on the regularization is to use the normalization of the $\delta$-like functions
$$
\lim_{\mu_{f},\mu_{g} \to 0}\int_{0}^{\infty}d|{\bf p}|\:\Bigl[\frac{\partial}{\partial |{\bf p}|}
{\tilde f}_{{\mu}_{f}}(|{\bf p}|)
\cdot{\tilde g}_{{\mu}_{g}}(|{\bf p}|)+{\tilde f}_{{\mu}_{f}}(|{\bf p}|)\cdot\frac{\partial}
{\partial |{\bf p}|}{\tilde g}_{{\mu}_{g}}(|{\bf p}|)\Bigr]=
$$
\begin{equation}
=\lim_{\mu_{f},\mu_{g} \to 0}\int_{0}^{\infty}d|{\bf p}|\:\frac{\partial}
{\partial |{\bf p}|}\Bigl[
{\tilde f}_{{\mu}_{f}}(|{\bf p}|)\cdot{\tilde g}_{{\mu}_{g}}(|{\bf p}|)\Bigr]
=\lim_{\mu_{f},\mu_{g} \to 0}\Bigl[
{\tilde f}_{{\mu}_{f}}(|{\bf p}|)\cdot{\tilde g}_{{\mu}_{g}}(|{\bf
p}|)\Bigr]^{\infty}_{0}
=-1
\end{equation}
and to respect the antisymmetry of the Schwinger term which implies the equality
of the first 2 terms (see Eq. (\ref{5.2})).

So the antisymmetry of the Schwinger term restricts already the general
possibilities for regularization and we are led to the following theorem.

\newtheorem{axiom}{Theorem}
\begin{axiom}
The classes of $\delta$-like functions $\{f_{{\mu}_{f}}(|{\bf x-y}|)\}$ and
$\{g_{{\mu}_{g}}(|{\bf x-y}|)\}$ which satisfy the integral conditions
\begin{equation}
\lim_{\mu_{f},\mu_{g} \to 0}\int_{0}^{\infty}d|{\bf p}|\:\frac{\partial}{\partial |{\bf p}|}{\tilde f}_{{\mu}_{f}}(|{\bf p}|)
\cdot{\tilde g}_{{\mu}_{g}}(|{\bf p}|)=-\frac{1}{2} \label{5.6},
\end{equation}
\begin{equation}
\lim_{\mu_{f},\mu_{g} \to 0}\int_{0}^{\infty}d|{\bf p}|\:\frac{\partial}
{\partial |{\bf p}|}{\tilde g}_{{\mu}_{g}}(|{\bf p}|)
\cdot{\tilde f}_{{\mu}_{f}}(|{\bf p}|)=-\frac{1}{2} \label{5.7},
\end{equation}
where both limits are of the same type,
will provide a {\rm unique} result for the Schwinger term $S(f,g)$.
\end{axiom}
This is the above mentioned integral condition on the classes of regularization
kernels {\sl and} it also gives a condition on how $\mu_{f}$
and $\mu_{g}$ have to approach zero.
For example, in the above described Gaussian or power kernel case the integral
condition is satisfied for the value $\mu=1$, which is actually the most natural
regularization, whereas in a combination of Gaussian and power kernels we must
choose a special value of $\xi$ depending on the value of $b$. Theorem 1 gives
us the possibility to use every combination of regularization kernels and to
define how $\mu_{f}$ and $\mu_{g}$ have to approach zero.

Finally, arriving such at a unique result the Schwinger term of the Gauss-law commutator is given by
$$
S^{ab}({\bf x-y})=\frac{-i}{24\pi^{2}}
\varepsilon^{ijk}\times\;\;\;\;\;\;\;\;\;\;\;\;\;\;\;\;\;\;\;\;\;\;\;\;\;\;\;\;\;\;\;\;\;\;\;\;\;\;\;\;\;\;\;\;\;\;\;\;\;\;\;\;\;
\;\;\;\;\;\;\;\;\;\;\;\;\;\;\;\;\;\;\;\;\;\;\;\;\;\;\;\;\;\;\;\;
$$
\begin{equation}
\times\;\mbox{tr}\Big{\{}
(\partial_{j}A_{k}+A_{j}A_{k})\{T^{a},T^{b}\}\partial_{i}\delta^{3}({\bf x-y})+
\frac{1}{2}\partial_{i}(A_{j}A_{k})\delta^{3}({\bf x-y})
[T^{a},T^{b}]\Big{\}}.
\end{equation}
Note that precisely the terms proportional to $A_{j}A_{k}$ break Faddeev's
cohomological result, Eq. (\ref{1.6}) (as found by Jo \cite{11}).

\section{Conclusion}
When working with a generalized point-splitting method for the calculation
of the Schwinger term in the commutator of Gauss-law operators the occuring
ambiguity due to the choice of regularization kernels can be overcome.
The asymmetry of the Schwinger term restricts the possibilities for
regularization allowing such that classes of regularization kernels which
satisfy the integral conditions (\ref{5.6}) and (\ref{5.7}) lead to a unique result.
A result, however, which differs from Faddeev's cohomology solution
(\ref{1.6}).

\section*{Note added in proof}
When calculating the commutator of the Gauss law operator and the
Hamiltonian $[{\cal G},H]$, or equivalently the time derivative of the Gauss
law operator -- as suggested by the referee -- we should obtain the anomaly
in the divergence of the chiral current \cite{26}. However, the generalized
point-splitting method used here does not work and must be altered. This we
will present in a forthcoming publication \cite{27}.

\section*{Acknowledgement}
One of us (T.S.) is grateful to G. Kelnhofer for helpful discussions and to
\mbox {W. Porod}, P. Stockinger and T. W\"ohrmann for computer help.

He is also grateful for obtaining a scholarship from the Aktion \"Osterreich
-- Tschechische Republik.

\small\normalsize
\newpage

\begin{table}
\caption{The combination of the Gaussian and power kernels. Values of $\xi$ which show us,
 how the limit procedures must be done, for  different values of $b$, to satisfy integral conditions
(\ref{5.6}) and
(\ref{5.7}).}\label{tab1}


\begin{tabular}{rlllll}
$ b$ & 3& 4& 5& 6& 7\\ \hline
$\xi$ & 1.07748 & 2.04837 & 3.03504 & 4.02744 & 5.02194\\
\end{tabular}

\end{table}

\end{document}

%% file: psbox.tex
\def\temp{1.34}%
\let\tempp=\relax
\expandafter\ifx\csname psboxversion\endcsname\relax
  \message{PSBOX(\temp) loading}%
\else
    \ifdim\temp cm>\psboxversion cm
      \message{PSBOX(\temp) loading}%
    \else
      \message{PSBOX(\psboxversion) is already loaded: I won't load
        PSBOX(\temp)!}%
      \let\temp=\psboxversion
      \let\tempp= 
    \fi
\fi
\tempp
\let\psboxversion=\temp
\catcode`\@=11
%
%
\def\psfortextures{
\def\PSspeci@l##1##2{%
\special{illustration ##1\space scaled ##2}%
}}%
\def\psfordvitops{
\def\PSspeci@l##1##2{%
\special{dvitops: import ##1\space \the\drawingwd \the\drawinght}%
}}%
\def\psfordvips{
\def\PSspeci@l##1##2{%
\d@my=0.1bp \d@mx=\drawingwd \divide\d@mx by\d@my
\special{PSfile=##1\space llx=\psllx\space lly=\pslly\space%
urx=\psurx\space ury=\psury\space rwi=\number\d@mx
}}}%
\def\psforoztex{
\def\PSspeci@l##1##2{%
\special{##1 \space
      ##2 1000 div dup scale
      \number-\psllx\space \number-\pslly\space translate
}}}%
\def\psfordvitps{
\def\psdimt@n@sp##1{\d@mx=##1\relax\edef\psn@sp{\number\d@mx}}
\def\PSspeci@l##1##2{%
\special{dvitps: Include0 "psfig.psr"}
\psdimt@n@sp{\drawingwd}
\special{dvitps: Literal "\psn@sp\space"}
\psdimt@n@sp{\drawinght}
\special{dvitps: Literal "\psn@sp\space"}
\psdimt@n@sp{\psllx bp}
\special{dvitps: Literal "\psn@sp\space"}
\psdimt@n@sp{\pslly bp}
\special{dvitps: Literal "\psn@sp\space"}
\psdimt@n@sp{\psurx bp}
\special{dvitps: Literal "\psn@sp\space"}
\psdimt@n@sp{\psury bp}
\special{dvitps: Literal "\psn@sp\space startTexFig\space"}
\special{dvitps: Include1 "##1"}
\special{dvitps: Literal "endTexFig\space"}
}}%
\def\psfordvialw{
\def\PSspeci@l##1##2{
\special{language "PostScript",
position = "bottom left",
literal "  \psllx\space \pslly\space translate
  ##2 1000 div dup scale
  -\psllx\space -\pslly\space translate",
include "##1"}
}}%
\def\psforptips{
\def\PSspeci@l##1##2{{
\d@mx=\psurx bp
\advance \d@mx by -\psllx bp
\divide \d@mx by 1000\multiply\d@mx by \xscale
\incm{\d@mx}
\let\tmpx\dimincm
\d@my=\psury bp
\advance \d@my by -\pslly bp
\divide \d@my by 1000\multiply\d@my by \xscale
\incm{\d@my}
\let\tmpy\dimincm
\d@mx=-\psllx bp
\divide \d@mx by 1000\multiply\d@mx by \xscale
\d@my=-\pslly bp
\divide \d@my by 1000\multiply\d@my by \xscale
\at(\d@mx;\d@my){\special{ps:##1 x=\tmpx, y=\tmpy}}
}}}%
\def\psonlyboxes{
\def\PSspeci@l##1##2{%
\at(0cm;0cm){\boxit{\vbox to\drawinght
  {\vss\hbox to\drawingwd{\at(0cm;0cm){\hbox{({\tt##1})}}\hss}}}}
}}%
\def\psloc@lerr#1{%
\let\savedPSspeci@l=\PSspeci@l%
\def\PSspeci@l##1##2{%
\at(0cm;0cm){\boxit{\vbox to\drawinght
  {\vss\hbox to\drawingwd{\at(0cm;0cm){\hbox{({\tt##1}) #1}}\hss}}}}
\let\PSspeci@l=\savedPSspeci@l
}}%
%
%
\newread\pst@mpin
\newdimen\drawinght\newdimen\drawingwd
\newdimen\psxoffset\newdimen\psyoffset
\newbox\drawingBox
\newcount\xscale \newcount\yscale \newdimen\pscm\pscm=1cm
\newdimen\d@mx \newdimen\d@my
\newdimen\pswdincr \newdimen\pshtincr
\let\ps@nnotation=\relax
{\catcode`\|=0 |catcode`|\=12 |catcode`|
|catcode`#=12 |catcode`*=14
|xdef|backslashother{\}*
|xdef|percentother{
|xdef|tildeother{~}*
|xdef|sharpother{#}*
}%
\def\R@moveMeaningHeader#1:->{}%
\def\uncatcode#1{%
\edef#1{\expandafter\R@moveMeaningHeader\meaning#1}}%
\def\execute#1{#1}
\def\psm@keother#1{\catcode`#112\relax}
\def\executeinspecs#1{%
\execute{\begingroup\let\do\psm@keother\dospecials\catcode`\^^M=9#1\endgroup}}%
\def\@mpty{}%
\def\matchexpin#1#2{
  \fi%
  \edef\tmpb{{#2}}%
  \expandafter\makem@tchtmp\tmpb%
  \edef\tmpa{#1}\edef\tmpb{#2}%
  \expandafter\expandafter\expandafter\m@tchtmp\expandafter\tmpa\tmpb\endm@tch%
  \if\match%
}%
\def\matchin#1#2{%
  \fi%
  \makem@tchtmp{#2}%
  \m@tchtmp#1#2\endm@tch%
  \if\match%
}%
\def\makem@tchtmp#1{\def\m@tchtmp##1#1##2\endm@tch{%
  \def\tmpa{##1}\def\tmpb{##2}\let\m@tchtmp=\relax%
  \ifx\tmpb\@mpty\def\match{YN}%
  \else\def\match{YY}\fi%
}}%
\def\incm#1{{\psxoffset=1cm\d@my=#1
 \d@mx=\d@my
  \divide\d@mx by \psxoffset
  \xdef\dimincm{\number\d@mx.}
  \advance\d@my by -\number\d@mx cm
  \multiply\d@my by 100
 \d@mx=\d@my
  \divide\d@mx by \psxoffset
  \edef\dimincm{\dimincm\number\d@mx}
  \advance\d@my by -\number\d@mx cm
  \multiply\d@my by 100
 \d@mx=\d@my
  \divide\d@mx by \psxoffset
  \xdef\dimincm{\dimincm\number\d@mx}
}}%
%
\newif\ifNotB@undingBox
\newhelp\PShelp{Proceed: you'll have a 5cm square blank box instead of
your graphics (Jean Orloff).}%
\def\s@tsize#1 #2 #3 #4\@ndsize{
  \def\psllx{#1}\def\pslly{#2}%
  \def\psurx{#3}\def\psury{#4}
  \ifx\psurx\@mpty\NotB@undingBoxtrue
  \else
    \drawinght=#4bp\advance\drawinght by-#2bp
    \drawingwd=#3bp\advance\drawingwd by-#1bp
  \fi
  }%
\def\sc@nBBline#1:#2\@ndBBline{\edef\p@rameter{#1}\edef\v@lue{#2}}%
\def\g@bblefirstblank#1#2:{\ifx#1 \else#1\fi#2}%
{\catcode`\%=12
\xdef\B@undingBox{
\def\ReadPSize#1{
 \readfilename#1\relax
 \let\PSfilename=\lastreadfilename
 \openin\pst@mpin=#1\relax
 \ifeof\pst@mpin \errhelp=\PShelp
   \errmessage{I haven't found your postscript file (\PSfilename)}%
   \psloc@lerr{was not found}%
   \s@tsize 0 0 142 142\@ndsize
   \closein\pst@mpin
 \else
   \if\matchexpin{\GlobalInputList}{, \lastreadfilename}%
   \else\xdef\GlobalInputList{\GlobalInputList, \lastreadfilename}%
     \immediate\write\psbj@inaux{\lastreadfilename,}%
   \fi%
   \loop
     \executeinspecs{\catcode`\ =10\global\read\pst@mpin to\n@xtline}%
     \ifeof\pst@mpin
       \errhelp=\PShelp
       \errmessage{(\PSfilename) is not an Encapsulated PostScript File:
           I could not find any \B@undingBox: line.}%
       \edef\v@lue{0 0 142 142:}%
       \psloc@lerr{is not an EPSFile}%
       \NotB@undingBoxfalse
     \else
       \expandafter\sc@nBBline\n@xtline:\@ndBBline
       \ifx\p@rameter\B@undingBox\NotB@undingBoxfalse
         \edef\t@mp{%
           \expandafter\g@bblefirstblank\v@lue\space\space\space}%
         \expandafter\s@tsize\t@mp\@ndsize
       \else\NotB@undingBoxtrue
       \fi
     \fi
   \ifNotB@undingBox\repeat
   \closein\pst@mpin
 \fi
\message{#1}%
}%
%
%
\def\psboxto(#1;#2)#3{\vbox{
   \ReadPSize{#3}%
   \divide\drawingwd by 1000
   \divide\drawinght by 1000
   \d@mx=#1
   \ifdim\d@mx=0pt\xscale=1000
         \else \xscale=\d@mx \divide \xscale by \drawingwd\fi
   \d@my=#2
   \ifdim\d@my=0pt\yscale=1000
         \else \yscale=\d@my \divide \yscale by \drawinght\fi
   \ifnum\yscale=1000
         \else\ifnum\xscale=1000\xscale=\yscale
                    \else\ifnum\yscale<\xscale\xscale=\yscale\fi
              \fi
   \fi
   \divide\pswdincr by 1000 \multiply\pswdincr by \xscale
   \divide\pshtincr by 1000 \multiply\pshtincr by \xscale
   \divide\psxoffset by1000 \multiply\psxoffset by\xscale
   \divide\psyoffset by1000 \multiply\psyoffset by\xscale
   \global\divide\pscm by 1000
   \global\multiply\pscm by\xscale
   \multiply\drawingwd by\xscale \multiply\drawinght by\xscale
   \ifdim\d@mx=0pt\d@mx=\drawingwd\fi
   \ifdim\d@my=0pt\d@my=\drawinght\fi
   \message{scaled \the\xscale}%
 \hbox to\d@mx{\hss\vbox to\d@my{\vss
   \global\setbox\drawingBox=\hbox to 0pt{\kern\psxoffset\vbox to 0pt{
      \kern-\psyoffset
      \PSspeci@l{\PSfilename}{\the\xscale}%
      \vss}\hss\ps@nnotation}%
   \advance\pswdincr by \drawingwd
   \advance\pshtincr by \drawinght
   \global\wd\drawingBox=\the\pswdincr
   \global\ht\drawingBox=\the\pshtincr
   \baselineskip=0pt
   \copy\drawingBox
 \vss}\hss}%
  \global\psxoffset=0pt
  \global\psyoffset=0pt
  \global\pswdincr=0pt
  \global\pshtincr=0pt 
  \global\pscm=1cm 
  \global\drawingwd=\drawingwd
  \global\drawinght=\drawinght
}}%
%
%
\def\psboxscaled#1#2{\vbox{
  \ReadPSize{#2}%
  \xscale=#1
  \message{scaled \the\xscale}%
  \advance\drawingwd by\pswdincr\advance\drawinght by\pshtincr
  \divide\pswdincr by 1000 \multiply\pswdincr by \xscale
  \divide\pshtincr by 1000 \multiply\pshtincr by \xscale
  \divide\psxoffset by1000 \multiply\psxoffset by\xscale
  \divide\psyoffset by1000 \multiply\psyoffset by\xscale
  \divide\drawingwd by1000 \multiply\drawingwd by\xscale
  \divide\drawinght by1000 \multiply\drawinght by\xscale
  \global\divide\pscm by 1000
  \global\multiply\pscm by\xscale
  \global\setbox\drawingBox=\hbox to 0pt{\kern\psxoffset\vbox to 0pt{
     \kern-\psyoffset
     \PSspeci@l{\PSfilename}{\the\xscale}%
     \vss}\hss\ps@nnotation}%
  \advance\pswdincr by \drawingwd
  \advance\pshtincr by \drawinght
  \global\wd\drawingBox=\the\pswdincr
  \global\ht\drawingBox=\the\pshtincr
  \baselineskip=0pt
  \copy\drawingBox
  \global\psxoffset=0pt
  \global\psyoffset=0pt
  \global\pswdincr=0pt
  \global\pshtincr=0pt 
  \global\pscm=1cm
  \global\drawingwd=\drawingwd
  \global\drawinght=\drawinght
}}%
%
\def\psbox#1{\psboxscaled{1000}{#1}}%
\newif\ifn@teof\n@teoftrue
\newif\ifc@ntrolline
\newif\ifmatch
\newread\j@insplitin
\newwrite\j@insplitout
\newwrite\psbj@inaux
\immediate\openout\psbj@inaux=psbjoin.aux
\immediate\write\psbj@inaux{\string\joinfiles}%
\immediate\write\psbj@inaux{\jobname,}%
%
%
\def\toother#1{\ifcat\relax#1\else\expandafter%
  \toother@ux\meaning#1\endtoother@ux\fi}%
\def\toother@ux#1 #2#3\endtoother@ux{\def\tmp{#3}%
  \ifx\tmp\@mpty\def\tmp{#2}\let\next=\relax%
  \else\def\next{\toother@ux#2#3\endtoother@ux}\fi%
\next}%
%
%
\let\readfilenamehook=\relax
\def\re@d{\expandafter\re@daux}
\def\re@daux{\futurelet\nextchar\stopre@dtest}%
\def\re@dnext{\xdef\lastreadfilename{\lastreadfilename\nextchar}%
  \afterassignment\re@d\let\nextchar}%
\def\stopre@d{\egroup\readfilenamehook}%
\def\stopre@dtest{%
  \ifcat\nextchar\relax\let\nextread\stopre@d
  \else
    \ifcat\nextchar\space\def\nextread{%
      \afterassignment\stopre@d\chardef\nextchar=`}%
    \else\let\nextread=\re@dnext
      \toother\nextchar
      \edef\nextchar{\tmp}%
    \fi
  \fi\nextread}%
\def\readfilename{\vbox\bgroup%
  \let\\=\backslashother \let\%=\percentother \let\~=\tildeother
  \let\#=\sharpother \xdef\lastreadfilename{}%
  \re@d}%
%
%
\xdef\GlobalInputList{\jobname}%
\def\psnewinput{%
  \def\readfilenamehook{
    \if\matchexpin{\GlobalInputList}{, \lastreadfilename}%
    \else\xdef\GlobalInputList{\GlobalInputList, \lastreadfilename}%
      \immediate\write\psbj@inaux{\lastreadfilename,}%
    \fi%
    \ps@ldinput\lastreadfilename\relax%
    \let\readfilenamehook=\relax%
  }\readfilename%
}%
\expandafter\ifx\csname @@input\endcsname\relax    
  \immediate\let\ps@ldinput=\input\def\input{\psnewinput}%
\else
  \immediate\let\ps@ldinput=\@@input
  \def\@@input{\psnewinput}%
\fi%
\def\nowarnopenout{%
 \def\warnopenout##1##2{%
   \readfilename##2\relax
   \message{\lastreadfilename}%
   \immediate\openout##1=\lastreadfilename\relax}}%
\def\warnopenout#1#2{%
 \readfilename#2\relax
 \def\t@mp{TrashMe,psbjoin.aux,psbjoint.tex,}\uncatcode\t@mp
 \if\matchexpin{\t@mp}{\lastreadfilename,}%
 \else
   \immediate\openin\pst@mpin=\lastreadfilename\relax
   \ifeof\pst@mpin
     \else
     \errhelp{If the content of this file is so precious to you, abort (ie
press x or e) and rename it before retrying.}%
     \errmessage{I'm just about to replace your file named \lastreadfilename}%
   \fi
   \immediate\closein\pst@mpin
 \fi
 \message{\lastreadfilename}%
 \immediate\openout#1=\lastreadfilename\relax}%
{\catcode`\%=12\catcode`\*=14
\gdef\splitfile#1{*
 \readfilename#1\relax
 \immediate\openin\j@insplitin=\lastreadfilename\relax
 \ifeof\j@insplitin
   \message{! I couldn't find and split \lastreadfilename!}*
 \else
   \immediate\openout\j@insplitout=TrashMe
   \message{< Splitting \lastreadfilename\space into}*
   \loop
     \ifeof\j@insplitin
       \immediate\closein\j@insplitin\n@teoffalse
     \else
       \n@teoftrue
       \executeinspecs{\global\read\j@insplitin to\spl@tinline\expandafter
         \ch@ckbeginnewfile\spl@tinline
       \ifc@ntrolline
       \else
         \toks0=\expandafter{\spl@tinline}*
         \immediate\write\j@insplitout{\the\toks0}*
       \fi
     \fi
   \ifn@teof\repeat
   \immediate\closeout\j@insplitout
 \fi\message{>}*
}*
\gdef\ch@ckbeginnewfile#1
 \def\t@mp{#1}*
 \ifx\@mpty\t@mp
   \def\t@mp{#3}*
   \ifx\@mpty\t@mp
     \global\c@ntrollinefalse
   \else
     \immediate\closeout\j@insplitout
     \warnopenout\j@insplitout{#2}*
     \global\c@ntrollinetrue
   \fi
 \else
   \global\c@ntrollinefalse
 \fi}*
\gdef\joinfiles#1\into#2{*
 \message{< Joining following files into}*
 \warnopenout\j@insplitout{#2}*
 \message{:}*
 {*
 \edef\w@##1{\immediate\write\j@insplitout{##1}}*
\w@{
\w@{
\w@{
\w@{
\w@{
\w@{
\w@{
\w@{
\w@{
\w@{
\w@{\string\input\space psbox.tex}*
\w@{\string\splitfile{\string\jobname}}*
\w@{\string\let\string\autojoin=\string\relax}*
}*
 \expandafter\tre@tfilelist#1, \endtre@t
 \immediate\closeout\j@insplitout
 \message{>}*
}*
\gdef\tre@tfilelist#1, #2\endtre@t{*
 \readfilename#1\relax
 \ifx\@mpty\lastreadfilename
 \else
   \immediate\openin\j@insplitin=\lastreadfilename\relax
   \ifeof\j@insplitin
     \errmessage{I couldn't find file \lastreadfilename}*
   \else
     \message{\lastreadfilename}*
     \immediate\write\j@insplitout{
     \executeinspecs{\global\read\j@insplitin to\oldj@ininline}*
     \loop
       \ifeof\j@insplitin\immediate\closein\j@insplitin\n@teoffalse
       \else\n@teoftrue
         \executeinspecs{\global\read\j@insplitin to\j@ininline}*
         \toks0=\expandafter{\oldj@ininline}*
         \let\oldj@ininline=\j@ininline
         \immediate\write\j@insplitout{\the\toks0}*
       \fi
     \ifn@teof
     \repeat
   \immediate\closein\j@insplitin
   \fi
   \tre@tfilelist#2, \endtre@t
 \fi}*
}%
\def\autojoin{%
 \immediate\write\psbj@inaux{\string\into{psbjoint.tex}}%
 \immediate\closeout\psbj@inaux
 \expandafter\joinfiles\GlobalInputList\into{psbjoint.tex}%
}%
%
%
%
\def\centinsert#1{\midinsert\line{\hss#1\hss}\endinsert}%
\def\psannotate#1#2{\vbox{%
  \def\ps@nnotation{#2\global\let\ps@nnotation=\relax}#1}}%
\def\pscaption#1#2{\vbox{%
   \setbox\drawingBox=#1
   \copy\drawingBox
   \vskip\baselineskip
   \vbox{\hsize=\wd\drawingBox\setbox0=\hbox{#2}%
     \ifdim\wd0>\hsize
       \noindent\unhbox0\tolerance=5000
    \else\centerline{\box0}%
    \fi
}}}%
%
\def\at(#1;#2)#3{\setbox0=\hbox{#3}\ht0=0pt\dp0=0pt
  \rlap{\kern#1\vbox to0pt{\kern-#2\box0\vss}}}%
%
\newdimen\gridht \newdimen\gridwd
\def\gridfill(#1;#2){%
  \setbox0=\hbox to 1\pscm
  {\vrule height1\pscm width.4pt\leaders\hrule\hfill}%
  \gridht=#1
  \divide\gridht by \ht0
  \multiply\gridht by \ht0
  \gridwd=#2
  \divide\gridwd by \wd0
  \multiply\gridwd by \wd0
  \advance \gridwd by \wd0
  \vbox to \gridht{\leaders\hbox to\gridwd{\leaders\box0\hfill}\vfill}}%
%
\def\fillinggrid{\at(0cm;0cm){\vbox{%
  \gridfill(\drawinght;\drawingwd)}}}%
%
%
\def\textleftof#1:{%
  \setbox1=#1
  \setbox0=\vbox\bgroup
    \advance\hsize by -\wd1 \advance\hsize by -2em}%
\def\textrightof#1:{%
  \setbox0=#1
  \setbox1=\vbox\bgroup
    \advance\hsize by -\wd0 \advance\hsize by -2em}%
\def\endtext{%
  \egroup
  \hbox to \hsize{\valign{\vfil##\vfil\cr%
\box0\cr%
\noalign{\hss}\box1\cr}}}%
%
\def\frameit#1#2#3{\hbox{\vrule width#1\vbox{%
  \hrule height#1\vskip#2\hbox{\hskip#2\vbox{#3}\hskip#2}%
        \vskip#2\hrule height#1}\vrule width#1}}%
\def\boxit#1{\frameit{0.4pt}{0pt}{#1}}%
\catcode`\@=12 
%
 \psfordvips   

%% file: art3.bbl
\begin{references}
\bibitem {1}R. Stora, Algebraic structure and topological origin of anomalies,
in: Recent progress in gauge theories, 1983 Carg\`ese Lectures,
H. Lehmann (ed.), NATO ASI series, Plenum Press, New York 1984.
\bibitem {2}B. Zumino, Chiral anomalies and differential geometry, in: Relativity,
 groups and topology II, 1983 Les Houches Lectures, B.S. DeWitt
 and R. Stora (eds.), North-Holland, Amsterdam 1984.
\bibitem {3}R. Bertlmann, Anomalies in quantum field theory, International Series
of Monographs on Physics 91, Clarendon -- Oxford University Press, 1996.
\bibitem {4}C. Adam, R.A. Bertlmann, P. Hofer, La Rivista del Nuovo Cimento, Vol. 16,
 No. 8, 1993.
\bibitem {5}J. Schwinger, Phys. Rev. Lett. 3 (1959) 296.
\bibitem {6}J. Mickelsson, Commun. Math. Phys. 97 (1985) 361.
\bibitem {7}L. Faddeev, Phys. Lett. 145B (1984) 81.
\bibitem {8}L. Faddeev, S. Shatashvili, Theor. Math. Phys. 60 (1984) 770.
\bibitem {9}R. Jackiw, Field theoretic investigations in current algebra, Topological
investigations of quantized gauge theories, in: Current algebra and ano\-malies,
S.B. Treiman, R. Jackiw, B. Zumino and E. Witten (eds.), p. 81 and p. 211,
World Scientific, Singapore 1985.
\bibitem {10}R. Jackiw, Diverse topics in theoretical and mathematical physics,
World  Scientific, Singapore 1995.
\bibitem {11}S.-G. Jo, Phys. Rev. D35 (1987) 3179.
\bibitem {12}S.-G. Jo, Nucl. Phys. B259 (1985) 616.
\bibitem {13}M. Kobayashi, A. Sugamoto, Phys. Lett. 159B (1985) 315.
\bibitem {14}M. Kobayashi, K. Seo, A. Sugamoto, Nucl. Phys. B273 (1986) 607.
\bibitem {15}A.J. Niemi, G.W. Semenoff, Phys. Rev. Lett. 55 (1985) 927.
\bibitem {16}A.J. Niemi, G.W. Semenoff, Phys. Rev. Lett. 56 (1986) 1019.
\bibitem {17}A.J. Niemi, G.W. Semenoff, Nucl. Phys. B276 (1986) 173.
\bibitem {18}H. Sonoda, Phys. Lett. 156B (1985) 220.
\bibitem {19}H. Sonoda, Nucl. Phys. B266 (1986) 410.
\bibitem {20}S. Hosono, Nucl. Phys. B300 [FS 22] (1988) 238.
\bibitem {21}E. Langmann, J. Mickelsson, Phys. Lett. B338 (1994) 241.
\bibitem {22}L. Faddeev, S. Shatashvili, Phys. Lett. 167B (1986) 225.
\bibitem {23}V.S. Vladimirov, Uravnjenija mat. fiziki, Izd. Nauka, Moskva 1989.
\bibitem {24}I.S. Gradshteyn, I.M. Ryzhik, Table of integrals, series, and
products, Academic Press, London -- New York, 1965.
\bibitem {25}S. Wolfram, Mathematica, Addison -- Wesley Publishing Company, Inc. 1991.
\bibitem {26}K. Fujikawa, Phys. Lett. B171 (1986) 424.
\bibitem {27}C. Adam, R.A. Bertlmann, T. S\'ykora, to be published.
\end{references}
